\documentclass[3p,times,twocolumn]{elsarticle}

\usepackage{ecrc}

\volume{00}

\firstpage{1}

\journalname{Nuclear Physics B Proceedings Supplement}

\runauth{M. Kordell, A. Majumder}

\jid{nppp}

\jnltitlelogo{Nuclear Physics B Proceedings Supplement}

\usepackage{amssymb}

\usepackage[figuresright]{rotating}

\usepackage{subcaption}

\begin{document}

\begin{frontmatter}



\dochead{}

\title{Jets and Centrality in p(d)-A Collisions}


\author{Michael C. Kordell II}
\address{Wayne State University, Detroit, MI 48202, USA}
\author{Abhijit Majumder}
\address{Wayne State University, Detroit, MI 48202, USA}

\begin{abstract}
The production of jets, and high-$p_{T}$ leading pions from jets, in d-Au collisions at the Relativistic Heavy-Ion Collider (RHIC) and p-Pb collisions at the Large Hadron Collider (LHC) are studied. Using a modified version of the event generator PYTHIA, in conjunction with a nuclear Glauber Monte-Carlo event generator, we demonstrate how events with a hard jet may be simulated, in such a way that the parton distribution function of the projectile nucleon is frozen during its interaction with the extended nucleus. Using our hybrid Monte-Carlo event generator, we demonstrate that the enhancement in $R_{pA}$ seen in peripheral events at RHIC and at LHC, as well as the depletion in central or semi-central events, is mainly due to ``mis-binning'' of central and semi-central events with a jet, as peripheral events. This occurs due to the reduction of soft particle production caused by a depletion of energy available in a nucleon (of the deuteron in the case of d-Au collisions), after the production of a hard jet. This represents a form of ``color transparency" of the projectile nucleon, which has fluctuated to a state with fewer and harder partons, in events which lead to jet production. We conclude with discussions of the form of multi-parton correlations in a nucleon which may be responsible for such a startling effect.
\end{abstract}


\end{frontmatter}


\section{Introduction}

Experimental measurements of $d-Au$ collisions at the Relativistic Heavy Ion Collider (RHIC) and $p-Pb$ collisions at the Large Hadron Collider (LHC) have shown a peculiar centrality dependent effect on the nuclear modification factor ($R_{p/d-A}$) for jets and leading hadrons as shown in Ref. \cite{Perepelitsa:2015aa}.  There is an observed enhancement in peripheral events and a result consistent with unity or further suppressed for central events as seen in Refs. \cite{Sahlmueller:2012ru}, \cite{ATLAS:2014cpa}, and \cite{ATLAS:2014cza}.  This is in conflict with our naive expectation that peripheral events should behave as p-p events and that any deviation from unity would be more pronounced in central events due to nuclear effects.

We present an explanation for these results as well as a test of the effects of this explanation on $R_{p/d-A}$ for comparison to these results.  In a given $p/d-A$ event where there is a jet, the nucleon's parton distribution function (PDF) must be in a state where there is at least one hard parton.  However, when in this state, there is a reduced amount of energy available to produce soft partons.  This depletion of soft partons results in a lower number of soft particles being produced for centrality binning.  In addition, at these ultra-relativistic energies, the PDF cannot change (fluctuate) between subsequent nucleon-nucleon collisions.  This produces a net effect that, in an event with a jet, lower numbers of soft particles are produced than what would have been produced if the event did not have a jet.  This causes the event to be binned as a more peripheral event than it would had otherwise been.

To study this, we setup a Monte-Carlo event generator for $p-Pb$ and $d-Au$ events.  The program starts by generating a heavier nucleus ($Pb$ or $Au$) and deuteron for $d-Au$ events.  These are handed off to an impact monte carlo that determines preliminary event information to be fed into PYTHIA.  In order to properly use PYTHIA for this analysis, PYTHIA had to be modified in order to incorporate event-by-event shadowing, in addition to a modification to prevent PDF resampling during the event.  PYTHIA observables were then used to determine $R_{pPb}$ or $R_{dAu}$ respectively to compare to experiment.  In the remainder of these proceedings, we will describe each of these components.  We will conclude with a discussion of the results from our simulations and an outlook for future work.

\section{Monte-Carlo Event Generation}

The nuclear Monte-Carlo was used to generate nuclei based upon an input density distribution.  To generate heavy nuclei for our analysis ($Au$ and $Pb$) we used a Woods-Saxon distribution
\begin{equation} \rho = \frac{1}{1 + \exp\left[ \frac{r - R}{a} \right ] }. \end{equation}
for $Au$, the parameters used were a = 0.535 fm, R = 6.38 fm and for $Pb$ the parameters used were a = 0.546 fm, R = 6.62 fm) and for the lighter deuteron we used a distribution based upon the Hulth\'{e}n wavefunction
\begin{equation} \psi = \frac{e^{-ar} - e^{-br}}{r}. \end{equation}
with parameters a = 0.228 $fm^{-1}$, b = 1.18 $fm^{-1}$.   The parameters and functions used were obtained from Ref. \cite{Greiner:1996aa}.  To account for the short-distance nucleon repulsion force, an exculsion volume for subsequently generated nucleons was used; this was set to be the diameter of a proton, 1.12 fm.  The Woods-Saxon distribution was sampled out to 13.3 fm and the Hulth\'{e}n distribution was sampled out to 38.9 fm to ensure that the rare fluctuations of nucleons well outside the central dense nucleus were accounted for.  The large value for the Hulth\'{e}n distribution was used to ensure that the long tails of distribution were captured with an error less than $3.67 \cdot 10^{-11}$.

The generated nuclei were then fed into the impact Monte Carlo.  The nuclei are projected onto the plane perpendicular to their momentum and an impact parameter was chosen for the event.  This was performed by sampling an $r^{2}$ distribution out to 10 fm; distances greater than this required more than ten million generated events to provide even a single collision.   In addition, a random $\theta$ was chosen from 0 to 2$\pi$ to allow the colliding nuclei to shift in angle relative to one another.  The second (lighter) nucleus was adjusted in position by the impact parameter and nucleon-nucleon collisions were counted by determining how many nucleons in the heavier nucleus were within a distance of the diameter of a proton from the colliding nucleon.  If this particular sampling did not result in at least a single nucleon-nucleon collision, the Monte-Carlo would reset and choose a new impact parameter, repeating the previous procedure until a collision was generated.  At this stage, relative probabilities of proton and neutron participation were evaluated, and the nucleon-nucleon collision types were determined (eg. $p-n$, $p-p$, etc.).

In order to generate events and event observables, the PYTHIA event generator was used \cite{Sjostrand:2006za}.  While it is not natively capable of full heavy-ion events, it is possible to modify PYTHIA to suit the purposes of this analysis.  These modifications to this program were made so that it could perform event-by-event nuclear shadowing, as well as alterations that would allow us to \textit{freeze} the PDF.

To implement nuclear shadowing, PYTHIA was altered to incorporate shadowing on an event-by-event basis.  This was accomplished by modifying the PDF calls with a factor of
\begin{equation} f(x) = 1 + (s(x) - 1) \cdot (N_{coll}/<N_{coll}>). \end{equation}
where $s(x)$ is the nuclear shadowing function from Ref. \cite{Li:2001xa}.  This will return the average value of the shadowing over a large number of events, but incorporates event-by-event fluctuations.  The number of collisions was used to adjust the shadowing.

To incorporate the fact that the PDF of an impacting nucleon does not resample during an event, including events where a hard parton was produced, the event was generated as a single impact of a proton on a "superproton".  This is in constrast to the more typical procedure where the event is treated as the sum of $n$ nucleon-nucleon collisions.  This was accomplished by taking the number of nucleons and multiplying the PDF calls.  By taking into account the number of neutrons and protons (determined earlier during the impact Monte Carlo stage) and multiplying the quark distribution calls within the PDF calls, preservation of isospin can be maintained.  As an example, the up quark distribution within the superproton will look like
\begin{equation} F^{S}_{u}(x) = N_{p} \cdot F_{u}(x) + N_{n} \cdot F_{d}(x). \end{equation}
for $N_{p}$ protons and $N_{n}$ neutrons and neglecting shadowing.
However, because PYTHIA treats the superproton as only having charge +1, net charge for the entire event will not be generated correctly.  At the energies of interest however, this does not effect the number of charged particles generated.

\section{Results}

\begin{figure}[htb]
	\includegraphics[width=0.4\textwidth]{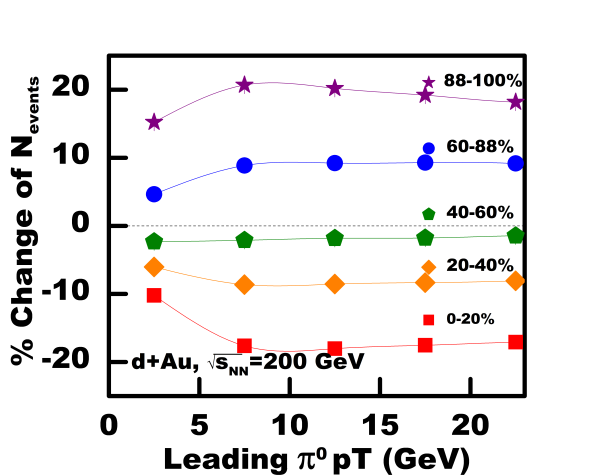}
	\caption{Percentage change of events in a given centrality bin due to "bin-shifting" as a function of $p_{T}$ for $d-Au$ events}
	\label{fig1}
\end{figure}

Before comparing to experiment we constructed a plot to quantify the effects of "bin-shifting"; where events in one centrality class as determined by $N_{bin}$ were shifted to another centrality class if binned by $N_{chg}$.  Generated events were tracked, and a plot of this bin-shifting was constructed as a function of the $p_{T}$ of the leading $\pi^{0}$ from the simulation for RHIC data as shown in Figure \ref{fig1}.

\begin{figure}[htb]
	\includegraphics[width=0.4\textwidth]{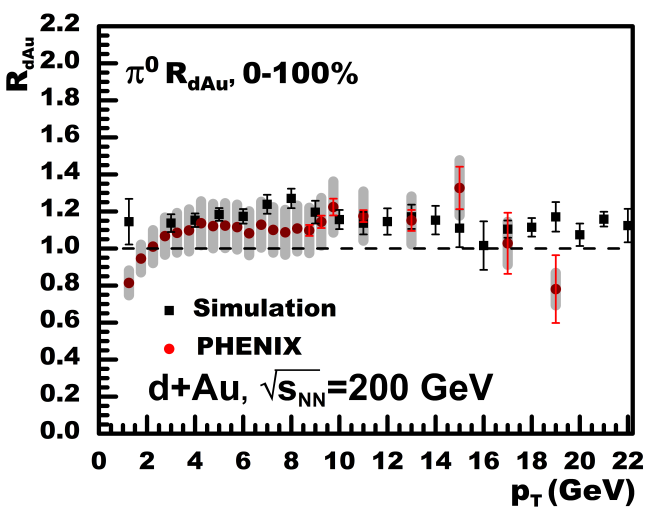}
	\caption{$R_{dAu}$ for Monte Carlo Event Generator for minimum bias compared to PHENIX results}
	\label{fig2}
\end{figure}

\begin{figure}[htb]
	\begin{subfigure}[b]{0.25\textwidth}
		\centering
		\includegraphics[width=\textwidth]{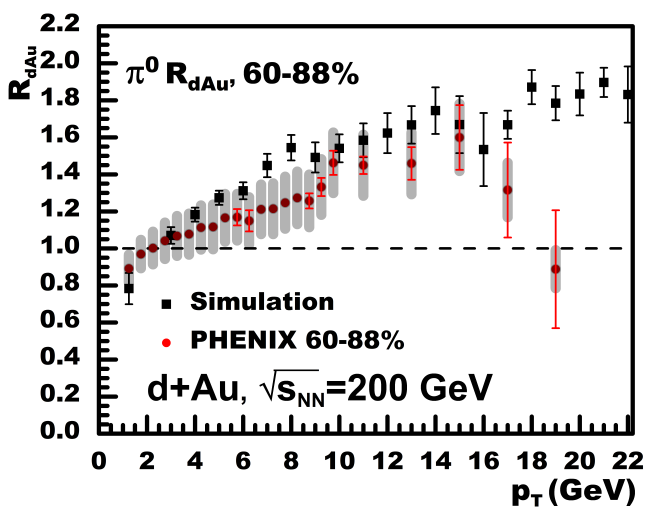}
	\end{subfigure}%
	\begin{subfigure}[b]{0.25\textwidth}
		\centering
		\includegraphics[width=\textwidth]{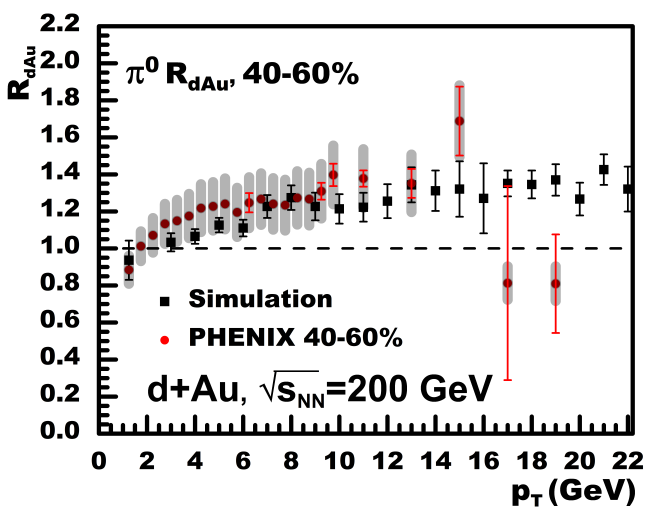}
	\end{subfigure}
	\begin{subfigure}[b]{0.25\textwidth}
		\centering
		\includegraphics[width=\textwidth]{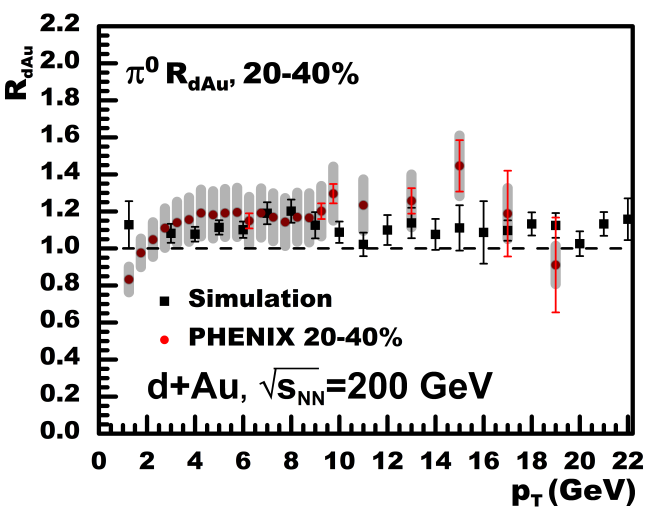}
	\end{subfigure}%
	\begin{subfigure}[b]{0.25\textwidth}
		\centering
		\includegraphics[width=\textwidth]{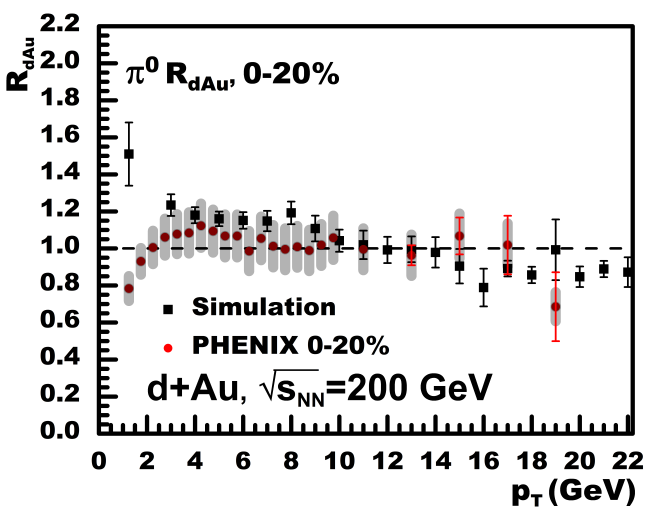}
	\end{subfigure}
	\caption{$R_{dAu}$ for Monte Carlo Event Generator with centrality determined by $N_{chg}$ compared to PHENIX results}
	\label{fig3}
\end{figure}

\begin{figure}[htb]
	\begin{subfigure}[b]{0.25\textwidth}
		\centering
		\includegraphics[width=\textwidth]{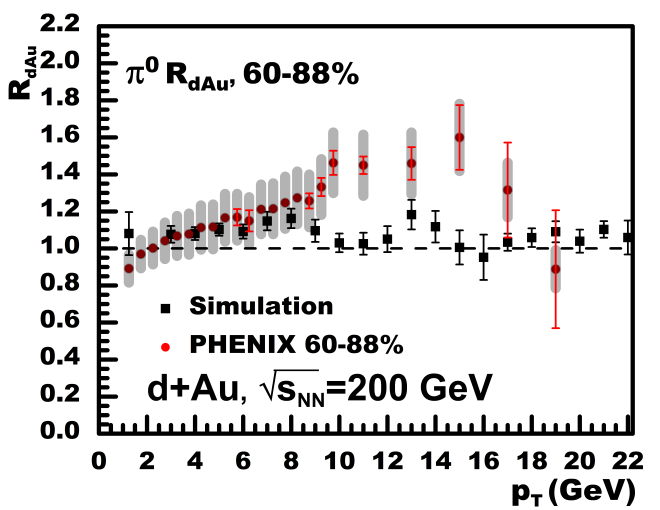}
	\end{subfigure}%
	\begin{subfigure}[b]{0.25\textwidth}
		\centering
		\includegraphics[width=\textwidth]{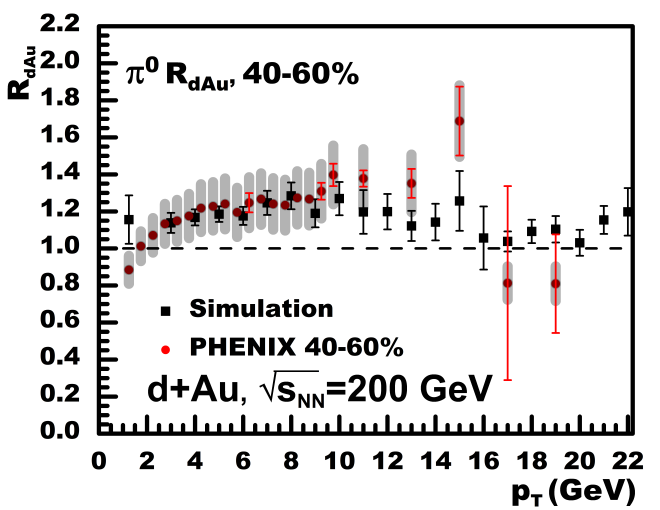}
	\end{subfigure}
	\begin{subfigure}[b]{0.25\textwidth}
		\centering
		\includegraphics[width=\textwidth]{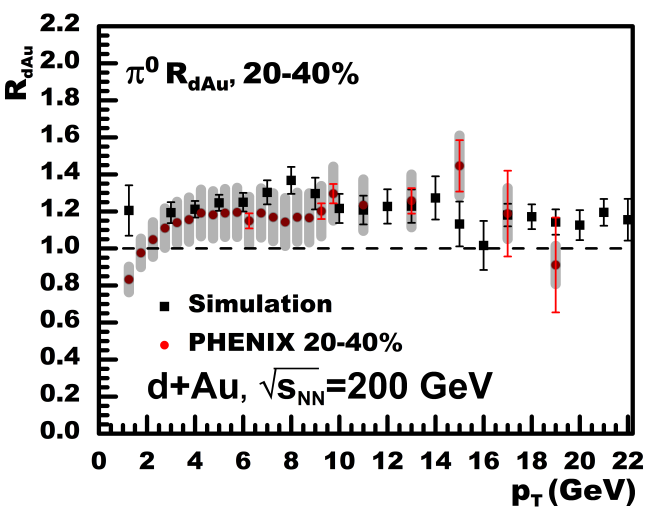}
	\end{subfigure}%
	\begin{subfigure}[b]{0.25\textwidth}
		\centering
		\includegraphics[width=\textwidth]{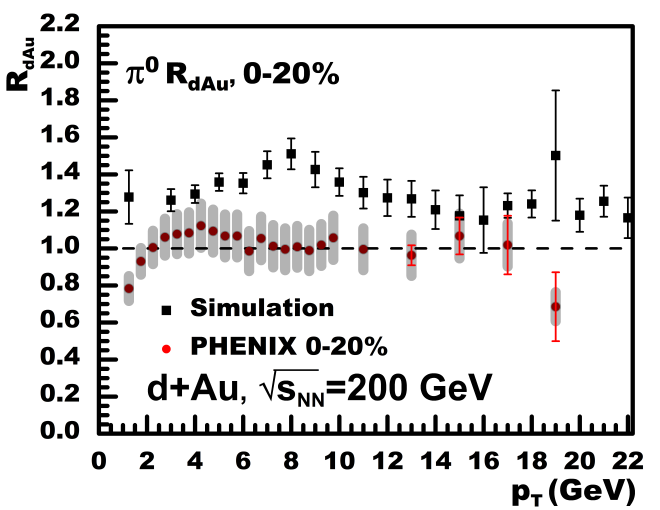}
	\end{subfigure}
	\caption{$R_{dAu}$ for Monte Carlo Event Generator with centrality determined by $N_{bin}$ compared to PHENIX results}
	\label{fig4}
\end{figure}

To compare to PHENIX $R_{dAu}$ data \cite{Sahlmueller:2012ru}, $\pi^{0}$s  at $y < |0.35|$ and $N_{chg}$ for each event were recorded.  While PHENIX determined centrality from $N_{chg}$ in the forward detector \cite{Adare:2013nff}, we determined centrality from $N_{chg}$ over the entire event.  This was necessary due to low statistics.  However this did not cause the minimum bias $R_{dAu}$ from generated events to differ from experiment.  For each centrality bin, $dN/dp_{T}dy$ was calculated.  Using a pion spectrum from pp events generated by PYTHIA, $R_{dAu}$ was calculated.  This is shown in Figure \ref{fig4} and a comparison plot for $R_{dAu}$ where centrality was determined with $N_{bin}$ is shown in Figure \ref{fig3}.  The minimum bias $R_{dAu}$ was also plotted to demonstrate the veracity of this event generator in the absence of any centrality dependent effects; this is shown in Figure \ref{fig2}.

~~

A similar procedure was performed for $p-Pb$ data from ATLAS, noting that there are two sets of data we wish to compare.  The first being $R_{CP}$ for inclusive jets \cite{ATLAS:2014cpa} and the second was $R_{CP}$ for charged hadron production \cite{ATLAS:2014cza}.   Experimentally, binning was performed by $E_{T}$ in the pseudorapidity interval $3.1 < \eta < 4.9$ in the direction of the lead beam \cite{ATLAS:2014cpa} for both data sets.  Our centrality binning was performed  by finding the total energy deposited outside of the two highest $p_{T}$ jets in the event or by collecting all charged particles over the entire event that carried less than 10 GeV of $p_{T}$.

\begin{figure}[htb]
	\centering
	\includegraphics[width=0.4\textwidth]{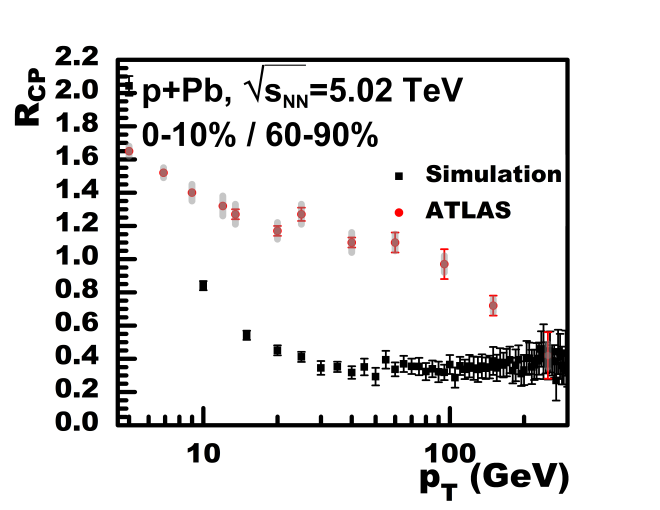}
	\caption{$R_{CP}$ from Monte Carlo Event Generator compared to ATLAS results for charged hadrons}
	\label{fig5}
\end{figure}

\begin{figure}[htb]
	\centering
	\includegraphics[width=0.4\textwidth]{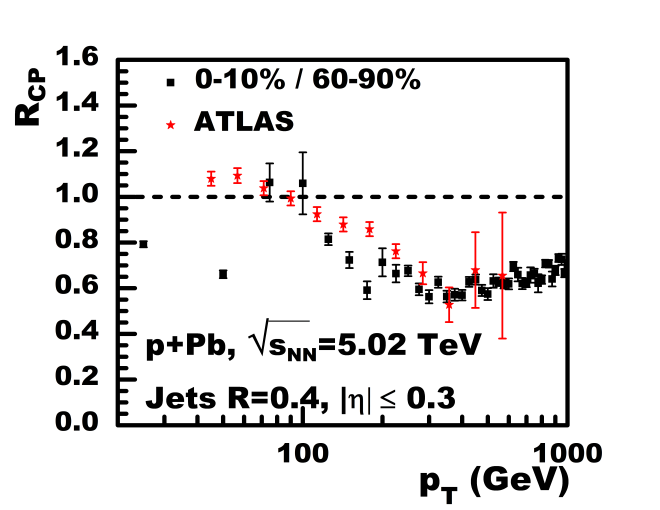}
	\caption{$R_{CP}$ from Monte Carlo Event Generator compared to ATLAS results for jets}
	\label{fig6}
\end{figure}

For jet $R_{CP}$, FastJet was used to determine jet properties for each event.  The anti-kT algorithm was used with R=0.4.  The two highest $p_{T}$ jets were recorded for each event, along with the total $E_{T}$ outside of the top two jets.  Using these jets, $dN/dp_{T}dy$ was calculated for the 0-10\% and 60-90\% centrality bins  which were divided to obtain $R_{CP}$.  The procedure was repeated for the charged hadron data, except centrality was determined by $N_{chg}$.  These are shown in Figures \ref{fig5} and \ref{fig6}.


\section{Conclusion}

In conclusion, we see that the centrality dependent effect on $R_{dAu}$ of this event generator when binned by $N_{chg}$ is consistent with PHENIX experimental results.  In addition, when binning by $N_{bin}$, we obtain the naive expectation that $R_{dAu}$ for peripheral events is consistent with unity and that for central events we see a slight enhancement due to nuclear shadowing.  For LHC p-Pb events, we find agreement between jet $R_{CP}$, however, there is a discrepancy for charged hadron $R_{CP}$.  This is likely due to the enhanced color connections in the superproton.

The core of this event generator relies on the fact that generation of a hard parton to create a jet results in the suppression of the production of soft particles; when a nucleon's PDF is in a state where there is a hard parton, it depletes the low x portion of the PDF.  Since binning of events is performed by some measure of soft particle production, this leads to events with jets being shifted to more peripheral bins than those events without a jet.  This was incorporated by "freezing" the PDF with PYTHIA; treating the event as a proton-superproton event invokes PYTHIA's energy conservation routines over the entire event.  With this event generator, we simulate the observed centrality dependant effects on $R_{p/d-A}$ and $R_{CP}$ for $d-Au$ and $p-Pb$ events as a result of the this "frozen PDF" effect.

This work was supported in part by the US National Science Foundation under grant number PHY-1207918 and by the US Department of Energy under grant number DE-SC00013460. This work is also supported in part by the Director, Office of Energy Research, Office of High Energy and  Nuclear Physics, Division of Nuclear Physics, of the U.S. Department of Energy, through the JET topical collaboration.





\end{document}